\documentclass{article}
\usepackage{epsf}
\title{On the Green's function and iterative solutions of Loop Quantum Cosmology}
\author{Fatimah Shojai \& Ali Shojai}
\date{}
\begin{document}
\maketitle
\begin{center}
Department of Physics, University of Tehran, North Kargar Avenue, Tehran 1439955961, Tehran, Iran;\\
and\\
Max-Planck-Institut f\"ur Gravitationsphysik, Alber-Einstein-Institut, Am M\"uhlenberg 1, D-14476 Potsdam, Germany.
\end{center}
\begin{abstract}
Here we shall find the green's function of the difference equation of loop quantum cosmology. To illustrate how to use it, we shall obtain an iterative solution for closed model and evaluate its corresponding Bohmian trajectory.
\\
Pacs: 04.60.Pp; 04.60.Ds; 03.65.Ta
\end{abstract}
\section{Introduction}
In loop quantum cosmology the evolution of the universe has been investigated from different points of view. Some of the results are based on the discrete quantum domain whose dynamics is described by a difference equation\cite{martin1,martin2}. As one expects the effects of discreteness are most important at small volumes, close to the classical singularity. In this limit the evolution of the universe is completely different from that given by the Wheeler-DeWitt equation. Therefore one has to learn how to extract results directly from the difference equation. The behaviour of quantum states near the Planck scale is not entirely known and only some numerical results exist. In \cite{ClosedExp} using an isotropic closed model with a massless scalar field, the divergent behaviour of the wave function at large scale is investigated numerically.
One of the most important consequences of discrete dynamics is the possibility of evolving from negative values of the scale factor eigenvalues to positive ones for all homogeneous models\cite{martin1}. In this way the quantum dynamics does not break at the classical singularity.

At large scales, one can approximate the difference equation with a differential equation, where the discreteness leads only to small corrections. 
This is an intermediate semiclassical phase in which the evolution equation takes a continuous form \cite{martin3}. There is another analysis called effective classical analysis. In this approximation one uses the effective matter Hamiltonian in which instead of inverse volume operator, it's eigenvalue is used \cite{martin3,Closed,EffectiveHam}. In this way one gets the effective classical equation of motion valid at large volumes, describing the motion of the position of the wave packet in coordinate time\cite{Time}. The most recent 
phenomenological applications of loop quantum cosmology are based on these equations\cite{martin3,Bojowald:QMC,tsm03,bounce_closed}. 
Inflationary scenarios without the  no--graceful exit problem\cite{martin3,Closed,EffectiveHam,Bojowald:QMC,tsm03}, new perspectives on initial 
conditions for standard inflation\cite{tsm03,martin_kevin_2} and resolution of the Big crunch problem in closed universe\cite{bounce_closed} are 
some of them.

In general the difference equation of loop quantum cosmology has solutions which have rapid change in their values and even in their sign when the volume changes slightly\cite{mm1}. The notion of \textit{pre-classicality}\cite{mm1} eliminates such unphysical solutions which have highly oscillatory behaviour for large scales. The physical solutions at large scales should be wave packets which their (WKB and/or Bohmian) trajectories are the classical ones. So far this condition on the solutions of loop quantum cosmology is done using the generating function techniques\cite{mm2,mm3,mm4}.

Recently\cite{shojai} we have obtained the exact solution of the Hamiltonian difference equation for the vacuum case. In this paper, we shall use that vacuum solutions, and find the green function of the difference equation. This is an important step in obtaining analytical solutions of the difference equation. Because using the green function one can obtain at least an iterative solution. We shall use this green function to obtain an iterative solution of closed universe in loop quantum cosmology in the presence of a massless scalar field. Finally we shall investigate the Bohmian trajectories corresponding to this solution, and see that these trajectories are compatible with the preclassicality condition.
\section{The Difference equation of cosmology}
Classically, the symmetry of cosmological minisuperspace is introduced by choosing an isotropic triad $E^a_i=p\delta^a_i$ and connection $A^i_a=c\delta^i_a$. The only physical parameters, $p$ and $c$ are related to the more familiar scale factor by the relations $|p|=a^2$ and $c=\frac{1}{2}(k-\gamma\dot{a})$ where $k$ is the curvature parameter and $\gamma$ is the Immirizi-Barbero factor.
Then, the quantum dynamics of states is obtained from the Hamiltonian constraint leading  to the following  difference equation\cite{ash,boj,tec}:
\[
(V_{\ell+5\ell_0}-V_{\ell+3\ell_0})e^{-2i\Gamma\ell_0}\psi(\phi,\ell+4\ell_0)
-\Omega(V_{\ell+\ell_0}-V_{\ell-\ell_0})\psi(\phi,\ell)+
\]
\begin{equation}
(V_{\ell-3\ell_0}-V_{\ell-5\ell_0})e^{2i\Gamma\ell_0}\psi(\phi,\ell-4\ell_0)=
-\frac{1}{3} 8\pi G\gamma^3\ell_0^3\ell_p^2\hat{C}_{m}(\ell,\phi)\psi(\phi , \ell)
\label{diff}
\end{equation}
in which $\Omega=2-4\ell_0^2\gamma^2\Gamma (\Gamma-1)$, $\hat{C}_{m}$ is the matter ($\phi$) Hamiltonian, and  $\ell_p$  is the Planck length. $\Gamma$ is the spin connection parameter defined as $\Gamma=k/2$.
$\psi(\phi,\ell)$ is the coefficient of the expansion
of the state in terms of \textit{spin network} states:
\begin{equation}
\left | \psi \right\rangle =\sum_\ell \psi(\phi,\ell)\left| \ell \right\rangle
\end{equation}
and $V_\ell=\left ( \frac{ \gamma |\ell|}{6}\right )^{3/2}\ell_p^3$ are
the eigenvalues of the volume operator. Finally $|\ell\rangle$ represents the spin network states for this simple minisuperspace, and it is simply the eigenstate of $\hat{p}$.

There is a contribution from the eigenvalues of inverse volume operator in the matter Hamiltonian. These are represented by $d_{j,\mu}(\ell)$:
\begin{equation}
\widehat{a^{-3}}|\ell\rangle=d_{j,\mu}(\ell)|\ell\rangle
\end{equation}
where $\mu\in (0,1)$ and the positive integer $j$ represent the quantum ambiguities arising from constructing quantum operator from the 
corresponding classical expression\cite{tt}. These eigenvalues are bounded at very small scale so the inverse volume operator is well defined 
at classical singularity. For large $j$ 
the behavior of $d_{j,\mu}(a)$  \cite{c}
shows a peak at $a^*=(\gamma\ell_p^2j/3)^{1/2}$ and then approaches to zero for all values of $\mu$. Therefore the scale $a^*$ 
determines the size of the scale factor below which the geometrical density is significantly different from it's classical form. 
For $a<<a^*$ the density approaches to zero according to  a power law behavior $d_{j,\mu}(a)\sim a^{\frac{3}{1-\mu}}$
and after passing a peak at $a^*$ it decays classically ($\sim a^{-3}$).

If one chooses the matter part to be a massless scalar field (This is the model we shall use to demonstrate the use of the green function obtained in the next section.):
\begin{equation}
\hat{C}_m=\frac{1}{2}\widehat{a^{-3}}p_\phi^2=-\frac{\hbar^2}{2}
\widehat{a^{-3}}\frac{d^2}{d\phi^2}
\end{equation}
and assume an oscillatory matter field dependence for the coefficient of the state vector to the matter field:
\begin{equation}
\psi(\phi,\ell)=\Phi(\ell)e^{i\tilde{\omega}\phi}
\end{equation}
where $\tilde{\omega}$ is a constant. So the right hand side of the difference equation(\ref{diff}) can be written as:
\begin{equation}
-\frac{8\pi G}{3}\gamma^3\ell_0^3\ell_p^2\hat{C}_m(\ell,\phi)\psi(\phi,\ell)=
-\ell_0^3\gamma^3\ell_p^6\omega^2d_{j,\mu}(\ell)\psi(\phi,\ell)
\end{equation}
where $\omega^2=\frac{8\pi G\hbar^2}{6\ell_p^4}\tilde{\omega}^2$. Using a new variable $F(\ell)$:
\begin{equation}
F(\ell)=\left ( V_{\ell+\ell_0}-V_{\ell-\ell_0}\right )e^{-i\ell\Gamma/2}\Phi(\ell)
\label{wave}
\end{equation}
the difference equation (\ref{diff}) can be written as:
\begin{equation}
F(\ell+4\ell_0)-\Omega F(\ell)+F(\ell-4\ell_0)=Q(\ell)
\label{diff1}
\end{equation}
where:
\begin{equation}
Q(\ell)=-\lambda\ell_p^3d_{j,\mu}(\ell)
\frac{F(\ell)}{V_{\ell+\ell_0}-V_{\ell-\ell_0}}
\end{equation}
where $\lambda$ is a dimensionless parameter defined as $\lambda=\ell_0^3\ell_p^3\gamma^3\omega^2$.
\section{Green function}
In this section we shall find the green function of the difference equation (\ref{diff1}). By the green function we mean a function satisfying the relation:
\begin{equation}
{\cal G}(\eta+4\ell_0)-\Omega {\cal G}(\eta)+{\cal G}(\eta-4\ell_0)=\delta_{\eta , 0}
\label{green}
\end{equation}
The general solution to the equation (\ref{diff1}) is then can be written as:
\begin{equation}
F(\ell)=F^{(0)}+\sum_{\ell '} {\cal G}(\ell-\ell ')Q(\ell ')
\end{equation}
In order to make the above summation meaningful one can assume that $G$ has support only on a countable subset of the real line. That is to say, one can construct an
infinite number of green functions each denoted as ${\cal G}^{(\eta_0)}(m)= {\cal G}(m\eta_0)$, where $m$ is an integer and $\eta_0$ is a real number. Different values of
$\eta_0$ leads a green function and a solution covering a subset of the domain of $\ell$.
\subsection{Closed universe}
To derive the green function we need the vacuum solutions. From reference \cite{shojai} we know that for the case of closed universe the vacuum solution is of the
form:
\begin{equation}
e^{\pm\beta\ell};\ \ \ \ \ \ \ \textit{with}\ \ \beta=\frac{1}{4\ell_0}\cosh^{-1}\frac{\Omega}{2}
\end{equation}
The green function is then can be suggested as (the boundary condition is chosen such that the green function goes to zero as $|\eta|\rightarrow\infty$):
\begin{equation}
{\cal G}(\eta)=\left \{
\begin{array}{lc}
ae^{\beta\eta}&\eta<0\\
b&\eta=0\\
ce^{-\beta\eta}&\eta>0
\end{array}
\right .
\end{equation}
Substituting this in the relation (\ref{green}) determines the constants as:
\begin{equation}
a=b=c=\frac{-1}{2\sinh 4\beta\ell_0}
\end{equation}
and the that $\eta$, as expected from the discussion above, should comes in steps of $\eta_0$ with $\eta_0\ge 4\ell_0$. 
The fact that $\eta_0\ge 4\ell_0$ would be clear if one substitutes the above solution in the relation (\ref{green}), for $\eta < 4\ell_0$ one encounters inconsistency for the values of $a$, $b$, and $c$.  Therefore $\eta$ could not be smaller than $4\ell_0$.
It is clear that changing $\eta_0$ in the range
$[4\ell_0,8\ell_0]$ would cover all the range of $|\ell|>4\ell_0$.

The green function would be simplified then to:
\begin{equation}
{\cal G}^{(\eta_0)}(n-m)=\frac{-e^{-\beta\eta_0|n-m|}}{2\sinh 4\beta\ell_0}
\end{equation}
The solution is thus:
\begin{equation}
F(m\eta_0)=F^{(0)}(m\eta_0)+\sum_n {\cal G}^{(\eta_0)}(n-m)Q(\eta_0 n)
\end{equation}
As it is noted previously, this solution works only for $|\ell|>4\ell_0$. For $|\ell|<4\ell_0$, one has to use the difference equation to get:
\begin{equation}
F(\pm\epsilon)=Q(4\ell_0\pm\epsilon)-F(8\ell_0\pm\epsilon)-\Omega F(4\ell_0\pm\epsilon)
\end{equation}
where $0<\epsilon<4\ell_0$.
\subsection{Flat universe}
For this case the vacuum solution is \cite{shojai}:
\begin{equation}
e^{i\alpha\ell}\ \ \ \ and\ \ \  \ell e^{i\alpha\ell}\ \ \ \ with\ \ \alpha=\frac{j\pi}{2\ell_0}
\end{equation}
where $j$ is an integer number. Again the green function is a linear combination of vacuum solutions. Putting the suggestion in the difference equation of green function, fixes
some of the constants and we are left with the green function:
\begin{equation}
{\cal G}^{(\eta_0)}(n-m)=e^{i\alpha\eta_0(n-m)}\left ( a+\frac{\eta_0|n-m|}{8\ell_0}\right )
\end{equation}
where $a$ is some undetermined constant. The appearance of a constant in the green function comes from the fact that we have not fix the boundary condition in the large $\ell$
limit. It only demanded that it should be oscillatory. It must be noted that again this green function can be used only for $|\ell|>4\ell_0$, and for $|\ell|<4\ell_0$
one should use the difference equation itself.
\section{Iterative solution}
Using the green functions derived in the previous section, one can obtain an iterative solution for the wave function. One can rewrite the equation for $F$ as follows
in terms of the wave function:
\begin{equation}
\Phi(m)=\Phi^{(0)}(m)+\frac{\lambda}{\Delta V(m)}\sum_n\tilde{\cal G}(n-m)S(n)\Phi(n)
\end{equation}
where we have dropped $\eta_0$ for simplicity and $\Delta V(m)=V_{m\eta_0+\ell_0}-V_{m\eta_0-\ell_0}$, $S(n)=\ell_p^3 d_{j,\mu}(n)$ and $\tilde{\cal G}=e^{-i\Gamma\eta_0(n-m)/2}{\cal G}$. This can be written in terms of infinite
dimensional matrices as:
\begin{equation}
\Delta V \Phi=\Delta V \Phi^{(0)}+\lambda \tilde{\cal G}\frac{S}{\Delta V}\Delta V \Phi
\end{equation}
This has the formal solution:
\begin{equation}
\Delta V \Phi=\frac{1}{1-\lambda \tilde{\cal G}\frac{S}{\Delta V}}\Delta V \Phi^{(0)}
\end{equation}
which can be expanded as:
\begin{equation}
\Phi(m)=\Phi^{(0)}(m)+\frac{\lambda}{\Delta V(m)}\sum_n\tilde{\cal G}(n-m)S(n)\Phi^{(0)}+\cdots
\end{equation}
In order to guarantee the convergence of this solution one should choose $\lambda$ as small as needed. This makes some limitation on $\omega$.

In the following we shall apply this general solution to the case of closed and flat universe and derive the wave function up to the first order of iteration.
\subsection{Closed universe}
For a closed universe the zeroth order solution is:
\begin{equation}
\Phi^{(0)}_{\pm}(m)=e^{im\Gamma\eta_0/2}\frac{e^{\pm m\beta\eta_0}}{\Delta V(m)}
\end{equation}
If one ignores the small terms like $e^{-2\beta\eta_0 n}$ appearing in the sum needed for calculating the first order of iteration, in comparison to 1, and using the
different behavior of $d_{j,\mu}$ in different regimes one can obtain the closed form:
\begin{equation}
\Phi^{(1)}_{\pm}(m)=e^{im\Gamma\eta_0/2}\frac{e^{\pm m\beta\eta_0}}{\Delta V(m)}\left \{ 1-\frac{\lambda}{2\sinh 4\beta\ell_0}\left (\frac{S(m)}{\Delta V(m)} \pm \Sigma (\pm
m)\right ) \right \}
\label{s1}
\end{equation}
where
\begin{equation}
\Sigma (m)= \left \{ 
\begin{array}{lc}
\frac{\Psi(1,|m+1|)}{3\ell_0\eta_0^2} & |m|>m^*\\
\frac{S(m^*)}{\Delta V(m^*)}+\frac{\Psi(1,m^*)}{3\ell_0\eta_0^2}+\frac{\sum_{|m+1|}^{m^*}n^{(2+\mu)/(2-2\mu)}}{3\ell_0\eta_0^{(-2-\mu)/(2-2\mu)}} & |m|<m^*
\end{array}
\right .
\end{equation}
and $\Psi(1,x)=d^2\textbf{ln}\Gamma(x)/dx^2$, in which $\Gamma$ is the Gamma function. In the above relation $m^*$ corresponds to $a^*$.
\subsection{Flat universe}
For this case, the iterative solution is not useful. Because in the summations, terms like $1/n$ appears and this diverges. This clearly does not mean that the solution does not exist in this case. The flat isotropic cosmology with a massless scalar field investigated by Ashtekar et. al. \cite{mm5}, recently. They used a semiclassical state as an initial state and evolve it backward in time and show that bigbang is replaced by big bounce. Our iterative method diverges for this model, and this means that the above results cannot be obtained from it unless one find a way to regularize the iteration.  
\section{Trajectories}
\begin{figure}[htb]
\begin{center}
\epsffile{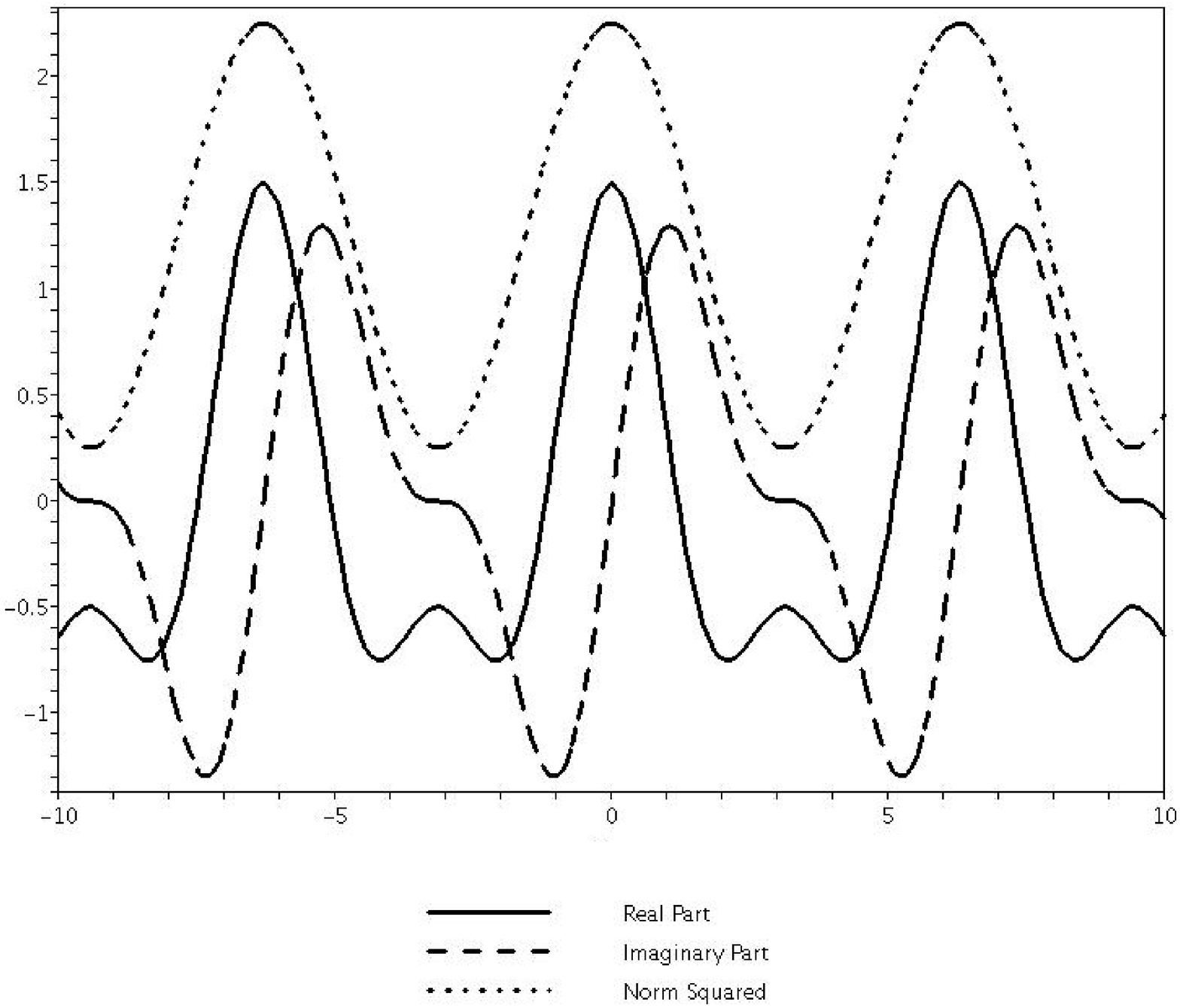}
\end{center}
\caption{The real part, imaginary part and norm squared of the wave function of a closed universe in the configuration space and up to first order of iteration. The horizontal axis is $\zeta c/2$, and the vertical axis is scaled by the factor: $\frac{1}{3\ell_0\sqrt{\zeta}}\left [ 1-\frac{\lambda}{2\sinh 4\beta\ell_0}\left ( \frac{\zeta^{3/2(1-\mu)}-S(m^*)}{3\ell_0\sqrt{\zeta}}-\frac{\Psi(1,m^*)}{3\ell_0\zeta^2}-\frac{\zeta^{(2+\mu)/2(1-\mu)}}{3\ell_0}\sum_{n=0}^{m^*}n^{\frac{(2+\mu)}{2(1-\mu)}}\right )\right ]$. In plotting the above, the oscillatory matter part $e^{i\tilde{\omega}\phi}$ is not included.}
\label{fig1}
\end{figure}

Here we shall find the Bohmian trajectories corresponding to the iterative wave function of closed universe obtained here. To do this, it is needed to have the wave function in the configuration space $|c\rangle$. Using the fact that $\langle c|\ell\rangle=e^{i\ell c/2}$, one has:
\begin{equation}
\psi(c,\phi)=\sum_{\eta_0}\sum_m\int d\omega f(\omega) g(\eta_0)e^{i\eta_0m(c+\Gamma)/2}e^{i\tilde{\omega}\phi}\Phi(m\eta_0)
\end{equation}
Here we choose $f(\omega)=\delta(\omega-\omega_0)$ and $g(\eta_0)=\delta_{\eta_0,\zeta}$ ($\zeta>4\ell_0$). Using the relation (\ref{s1}), one can obtain the wave function in the configuration space up to first order. It is plotted in figure (\ref{fig1}).
\epsfxsize=5in
\epsfysize=5in

\begin{figure}[htb]
\begin{center}
\epsffile{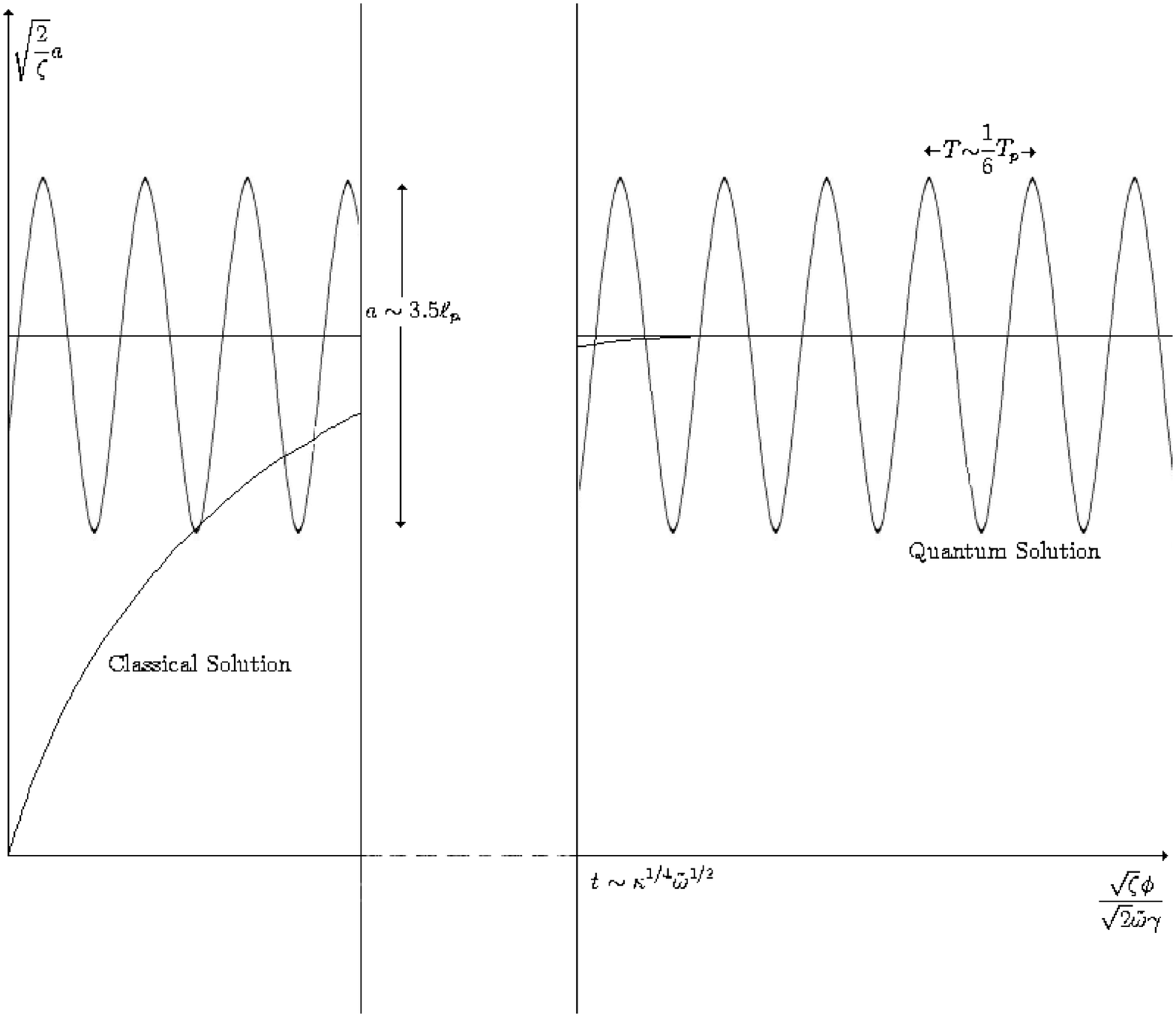}
\end{center}
\caption{The classical and quantum scale factors as a function of the matter field. The horizontal axis is $\frac{\sqrt{\zeta}\phi}{\sqrt{2}\tilde{\omega}\gamma}$ and the vertical axis is $\frac{\sqrt{2}a}{\sqrt{\zeta}}$. Note that for times greater than the classicality time ($\kappa^{1/4}\tilde{\omega}^{1/2}$), the quantum solution is oscillations of amplitude $\sim 3.5\ell_p$ with period $\sim \frac{1}{6}T_p$, where $T_p$ is planck's time. For small times, the Bohmian quantum trajectory has no singularity.}
\label{fig2}
\end{figure}

The corresponding Bohmian trajectory can be found simply by the guidance relation. Using relations of reference \cite{shojai}, we have:
\begin{equation}
p=\frac{\delta S}{\delta c}
\end{equation}
\begin{equation}
\dot{\phi}=\frac{\delta S}{\delta \phi}
\end{equation}
The coordinate time can be eliminated between these equations and $a$ as a function of $\phi$ derived. The result is plotted in figure (\ref{fig2}). 
Also in this figure the classical trajectory is plotted. 
The classical trajectory can be obtained from the classical Hamiltonian constraint\cite{ClosedExp}:
\begin{equation}
-\frac{3}{2\gamma^2\pi G}p^2\left ( c^2-c+\frac{1+\gamma^2}{4}\right )+p_\phi^2=0
\end{equation}
with $p_\phi=\tilde{\omega}$ one gets the Friedman equation:
\begin{equation}
\dot{a}^2+1=\frac{8\pi G\tilde{\omega}^2}{3a^4}
\end{equation}
which is independent of $\gamma$ as it should be. The solution to this equation is either $a=constant=(8\pi G\tilde{\omega}^2/3)^{1/4}$ or 
\begin{equation}
t=\int\frac{a^2da}{\sqrt{\frac{8\pi G \tilde{\omega}^2}{3}-a^4}}
\end{equation}
which can be expressed in terms of elliptic integrals.

It is important to note that the scale factor has oscillations of amplitude of order of $\ell_p$. For times far away from the classical singularity, the Bohmian trajectory is the classical one up to quantum oscillations (fluctuations) around it. This is in accordance with the pre-classicality condition.
\epsfxsize=5in
\epsfysize=5in

At this end it is fruitful to note some points. First that this is an iterative method based on the standard idea of Green's function. Naturally this may be of limited use. It is not possible to obtain the exact solution from the iterative one, in general. This method is usefull only if the iteration is convergent to the exact one.

Second, which is related to the first, is the reader may ask what is the meaning of smallness of $\lambda$? It is assumed that if one chooses $\lambda$ small enough, which means choosing the matter frequency $\omega$ small enough, it is possible to make the iteration convergent. Physically this means that if the matter frequency is small, the matter contribution to the difference equation would be small and one expects that iteration around the vacuum solution is convergent. But mathematically, one should examine whether each model is convergent by rigour mathematical methods.

Finally, it must be noted that the method presented here can, in principle, be extended to more complicated models. For example, for anisotropic models, the only difference is that we have a partial difference equation. The Green's function of this equation can be obtained via separation of variables. The result would be some kind of multiplication of the isotropic Green's function for each direction of the anisotropic model.
   
\textbf{Acknowledgment}: This work is supported by a grant from University of Tehran. The authors are grateful to Martin Bojowald for useful discusions and suggestions.


\begin{thebibliography}{99}
\bibitem{martin1}
M.~Bojowald, Phys. Rev. Lett., \textbf{86}, 5227, 2001, [arXiv:gr-qc/0102069].
\bibitem{martin2}
M.~Bojowald, Phys. Rev. Lett., \textbf{87}, 121301, 2001, [arXiv:gr-qc/0104072].
\bibitem{ClosedExp}
D. Green and W. Unruh, Phys. Rev. D, \textbf{70}, 103502, 2004, [arXive:gr-qc/0408074].
\bibitem{martin3}
M.~Bojowald, Phys. Rev. Lett., \textbf{89}, 261301, 2002, [arXiv:gr-qc/0206054].
\bibitem{Closed}
M. Bojowald and K. Vandersloot, Phys. Rev. D, \textbf{67}, 124023, 2003, [arXive:gr-qc/0303072].
\bibitem{EffectiveHam}
G.~Date and G.~M.~Hossain, Class. Quant. Grav., \textbf{21}, 4941, 2004, [arXiv:gr-qc/0407073].
\bibitem{Time}
M.~Bojowald, P.~Singh and A.~Skirzewski, Phys. Rev. D, \textbf{70}, 124022, 2004, [arXive:gr-qc/0408094].
\bibitem{Bojowald:QMC}
M.~Bojowald, J.~E.~Lidsey, D.~J.~Mulryne, P.~Singh and R.~Tavakol, Phys. Rev. D, \textbf{70}, 043530, 2004, [arXiv:gr-qc/0403106].
\bibitem{tsm03}
S.~Tsujikawa, P.~Singh \& R.~Maartens, Class. Quant. Grav., \textbf{21}, 5767, 2004, [arXiv:astro-ph/0311015].
\bibitem{bounce_closed}
P.~Singh \& A.~Toporensky, Phys. Rev. D, \textbf{69}, 104008, 2004, [arXiv:gr-qc/0312110];\\
G.~V.~Vereshchagin, JCAP, \textbf{0407}, 013, 2004, [arXiv:gr-qc/0406108].
\bibitem{martin_kevin_2}
M. Bojowald \& K. Vandersloot, Invited parallel talk at $X^{th}$ Marcel Grossmann meeting, July 20-26, Rio de Jaeiro, 2003, [arXiv:gr-qc/0312103].
\bibitem{mm1}
\textit{Living Rev. Relativity}, \textbf{8}, 11, 2005. Available online at: 
\begin{verbatim}
http://www.livingreviews.org/lrr-2005-11
\end{verbatim}
\bibitem{mm2}
D. Cartin, and G. Khanna, \textit{Phys. Rev. Lett.}, \textbf{94}, 111302, 2005.
\bibitem{mm3}
D. Cartin, G. Khanna, and M. Bojowald, \textit{Class. Quant. Grav.}, \textbf{21}, 4495, 2004.
\bibitem{mm4}
S. Connors, and G. Khanna, Lanl arXive:gr-qc/0509081.
\bibitem{shojai}
A. Shojai, and F. Shojai, \textit{EuroPhys. Lett.}, \textbf{71}, 6, 886, 2005.
\bibitem{ash}
A. Ashtekar, and J. Lewandowski, Class. Quant. Grav., \textbf{21}, R53, 2004.
\bibitem{boj}
A. Ashtekar, M. Bojowald, and J. Lewandowski, Adv. Theor. Math. Phys., \textbf{7}, 233, 2003.
\bibitem{tec} M. Bojowald, H.A. Morales-Tecotl, Lect. Notes Phys., \textbf{646},  421, 2004.
\bibitem{tt}T. Thiemann, Class. Quant. Grav., 15, 1281, 1998.
\bibitem{c}
M.~Bojowald, Pramana J. Phys., \textbf{63}, 765, 2004, [arXiv:gr-qc/0402053].
\bibitem{mm5}
A. Ashtekar, T. Pawlowski, and P. Singh, Lanl arXive:gr-qc/0602086.
\end{thebibliography}
\end{document}